\documentstyle[psfig]{mn}

\ifCUPmtlplainloaded \else
  
  \def\umu{\mu}
  
\fi

\newcommand{\um}{\ifmmode{\;\umu{\rm m}}\else$\;{\umu{\rm m}}$\fi}
\newcommand{\kms}{\ifmmode{\;{\rm km\,s^{-1}}}\else%
  ${\;{\rm km\,s^{-1}}}$\fi}
\newcommand{\ee}[1]{\ifmmode {} \times 10^{#1} \else ${} \times10^{#1}$\fi}
\newcommand{\ergs}{\ifmmode{\;{\rm erg\,s^{-1}}}\else%
  ${\;{\rm erg\,s^{-1}}}$\fi}
\newcommand{\ergscm}{\ifmmode{\;{\rm erg\,s^{-1}\,cm^{-2}}}
    \else${\;{\rm erg\,s^{-1}\,cm^{-2}}}$\fi}
\newcommand{\Msolar}{\ifmmode{\; {\rm M_{\sun}}}\else%
  {${\; {\rm M_{\sun}}}$}\fi}
\newcommand{\HeI}{He$\,${\sc i}}
\newcommand{\HI}{H$\,${\sc i}}

\newcommand{\eg}{{e.g.}}
\newcommand{\etal}{{et~al.}}
\newcommand{\ie}{{i.e.}}

\title[Spectroscopy of Cir X-1]{High resolution optical and infrared
  spectroscopic observations of Cir X-1} \author[Helen M. Johnston,
Robert Fender, and Kinwah Wu]{Helen M. Johnston,$^1$\thanks{E-mail:
    hmj@aaoepp.aao.gov.au}
  Robert Fender,$^2$\ and Kinwah Wu$^3$ \\
  $^1$ Anglo-Australian Observatory, P.O. Box 296, Epping NSW
  1710, Australia\\
  $^2$ Astronomical Institute `Anton Pannekoek', University of
  Amsterdam, Kruislaan 403, 
  1098 SJ Amsterdam, The Netherlands\\ 
  $^3$ Research Centre for Theoretical Astrophysics, School of
  Physics, University of Sydney, NSW 2006, Australia} \date{}

\pagerange{\pageref{firstpage}--\pageref{lastpage}}
\pubyear{1998}

\psfigurepath{/home/hmj/SCR1/Figures/CirX-1/}

\begin{document}

\maketitle

\label{firstpage}

\begin{abstract}
     We present new optical and infrared (IR) observations of Cir X-1
     taken near apastron. Both sets of spectra show asymmetric
     emission lines.  Archival optical observations show that an
     asymmetric H$\alpha$\ emission line has been in evidence for the
     past twenty years, although the shape of the line has changed
     significantly.  We present an eccentric ($e \sim 0.7$--0.9) low
     mass binary model, where the system consists of a neutron star
     orbiting around a (sub-)giant companion star of 3--5\Msolar. We
     suggest that the broad components of the emission lines arise in
     a high-velocity, optically thick flow near the neutron star;
     while the narrow components of the optical and the IR lines arise
     near the companion star and a heated ejecta-shell surrounding the
     binary respectively. In this model, the velocity of the narrow
     component reflects the space velocity of the binary; the implied
     radial velocity (+430\kms\ after correcting for Galactic rotation)
     is the highest velocity known for an X-ray binary.
\end{abstract}


\section{Introduction}
\label{sec:Introduction}

Cir~X-1 is one of the most puzzling X-ray binaries known. Like the
peculiar systems SS 433 and Cyg X-3, it cannot easily be classified
into any of the major categories of X-ray binaries. Indeed, there is
even doubt as to whether it is a high-mass (HMXB) or low-mass X-ray
binary (LMXB). 

Since its discovery in early 1970s, Cir~X-1 has been studied
intensively at X-ray wavelengths. The X-ray properties of Cir~X-1 were
found to differ dramatically each time it was observed (see for
example Kaluzienski et~al. 1976, Tennant 1988, Tsunemi et~al. 1989,
Shirey \etal\ 1996).\nocite{khbs76,ten88,tkm+89,sblm96} Periodic
modulation of the X-ray flux was found at a period of 16.6~d
\cite{khbs76}.  A radio counterpart was found \cite{cpc75}, and was
found to flare at the same period as the X-ray modulation
\cite{hjm+78}.  These flares were initially detected at peak flux
levels of $> 1\;$Jy; since the 1970s, the flux of the source has
decreased dramatically, and it has only occasionally been detected
above $50\;$mJy \cite{snp+91}.

This radio source is located $25'$\ from the centre of the supernova
remnant G321.9$-$0.3, and is apparently connected to the remnant by a
radio nebula \cite{hkl+86}. Stewart \etal\ \shortcite{schn93} have
imaged arcmin-scale collimated structures within the surrounding
nebula, suggesting an outflow from the X-ray binary.  Fender \etal\ 
\shortcite{fst+98} have imaged an arcsec-scale asymmetric jet aligned
with these larger structures, raising the possibility that the outflow
from the system is relativistic. Recently, Case \& Bhattacharya
\shortcite{cb98} have revised the estimated distance to G321.9$-$0.3
(and hence to Cir~X-1, assuming they are associated) to $5.5\;$kpc,
which is substantially smaller than the original suggested distance to
Cir~X-1 of $10\;$kpc \cite{gm77}.

The discovery of Type I X-ray bursts \cite{tfs86b} suggests that
the compact object is probably a weakly-magnetised neutron star.  The
close association of Cir~X-1 with the supernova remnant suggests that
the system may be a young ($< 10^5$~y old) runaway system from a
supernova explosion \cite{schn93}.

The optical counterpart to Cir~X-1 was identified as a highly-reddened
star with strong H$\alpha$\ emission \cite{wmw+77}.  This object
was later shown to consist of three stars within a radius of 1\farcs5,
the southernmost of which is the true counterpart
(\cite{mon92,dsh93}).

\begin{figure*}
\centerline{\psfig{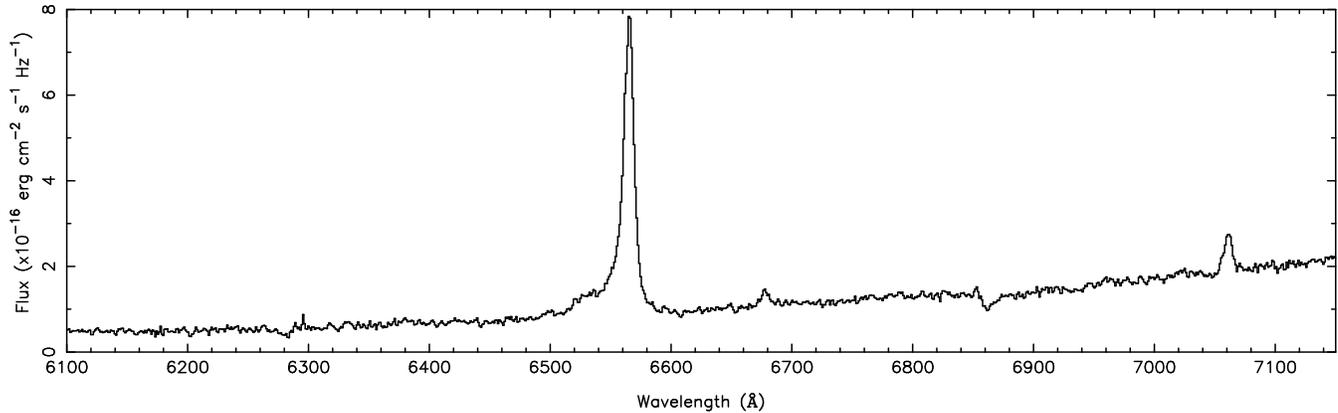}}
\caption{Spectrum of Cir X-1, taken near apastron on 1997 June 4: the
  spectrum shown is the sum of five 1800~s exposures. Emission lines
  of lines of H$\alpha$\ $\lambda$6563, \HeI\ $\lambda$6678, and
  \HeI\ $\lambda$7065 can be seen; the dip at $\lambda$6870 is
  due to imperfect removal of the terrestrial atmospheric
  B-band.}\label{fig:spectrum}
\end{figure*}

The long orbital period and periodic X-ray activity suggested a
high-mass system in an eccentric orbit \cite{mjh+80}; however,
the variability of the optical emission, the faintness of the optical
counterpart, and several of its X-ray characteristics suggest that the
companion is a low-mass star. The lack of spectroscopic studies in the
optical band means that most of the fundamental orbital parameters of
Cir~X-1 have not been determined. Moreover, Cir~X-1 shows very
different properties from time to time. This make it difficult to
construct a coherent picture for the system from observations of
different wavelengths at different epochs. Here we present new
spectroscopic observations of Cir~X-1, and use these, together with
analysis of archival observations of the system, to suggest a more
coherent model for the system.

\section{Observations and data reduction}
\label{sec:Observ-data-reduct}

\subsection{New optical observations}
\label{sec:New-optic-observ}

Cir~X-1 was observed on 1997 June 4 using the 3.9~m Anglo-Australian
Telescope (AAT). The mean orbital phase of the observation was 0.51,
calculated according the ephemeris of Stewart \etal\ 
\shortcite{snp+91}.  The RGO Spectrograph was used in combination with
the TEK 1k CCD in the 82~cm camera and a grating of 270
grooves$\;\mathrm{mm}^{-1}$\ in first order, resulting in a dispersion
of $\sim 1.08\;\mathrm{\AA}\;\mathrm{pixel}^{-1}$\ over a wavelength
range 6060--7165$\;$\AA. The spatial scale was $0\farcs25$; the
spectral resolution, measured from the arc lines, was
$5.4\;\mathrm{\AA}$.

A $1\farcs5$-wide slit was used, oriented north-south so both Cir~X-1
and star 2 of Moneti \shortcite{mon92} were in the slit. The
atmospheric seeing was about $1''$. Five 1800$\;$s integrations were
taken, interspersed with CuAr arc-lamp exposures, before cloud
prevented the acquisition of any more data.

The bias and pixel-to-pixel gain variations were removed from each
exposure using standard procedures in {\sc iraf}.  Cosmic rays were
removed using the method of Croke \shortcite{cro95} to compare
adjacent frames.  Because of the presence of the nearby confusing star
(Moneti's star 2), special care needed to be taken to measure the flux
from our object.  At every position along the dispersion direction, we
fit two gaussian profiles, with fixed widths (FWHM=5.6~pixels) and
separation (6 pixels), to the sky-subtracted frames.  The amplitude of
these gaussians was used as the estimate of the flux from Cir~X-1 and
star~2 at each wavelength. We then determined the wavelength
calibration using the CuAr arc lamp exposures. We fit a low-order
polynomial to the arc line wavelengths as a function of pixel number:
the rms scatter of the fits was $\sim 1/4$\ of a pixel. A rough
flux-calibration was performed by comparing with the spectrum of the
observed flux standard LTT~4364, though since the night was
non-photometric, this flux calibration should be considered only
approximate.

\subsection{Infrared observations}
\label{sec:Infr-observ}

$K$-band spectroscopy of Cir~X-1 was obtained using the Cryogenic
Array Spectrometer/Imager (CASPIR) on the ANU 2.3~m telescope at
Siding Spring Observatory on the night of 1997 June 20. The $K$\ grism
was used with the SBRC $256\times256$\ InSb array, giving a dispersion
of $21\farcs5$~\AA$\,{\mathrm pixel}^{-1}$\ over a wavelength range of
1.94--2.49\um. The spatial scale was $0\farcs5\;{\mathrm pixel}^{-1}$.
A $5''$\ slit was used, oriented east-west: note that this means that
Moneti's stars 2 and 3 both contributed light in our spectrum.  The
telescope was nodded by $\pm\, 12''$\ along the slit to provide sky
frames at the same position as the object.  Argon lamp spectra were
taken to perform wavelength calibration, and two nearby bright stars
(BS~5699 and BS 5712) were observed in order to remove atmospheric
spectral features and perform flux calibration.

Standard data reduction procedures were followed, using the local {\tt
  caspir} package running in {\sc iraf}.  Bias and dark frames were
used to linearise all frames, the sky background was subtracted from
the object frames, and pixel-to-pixel variations were corrected. The
chip distortion was corrected in order to align the dispersion and
spatial directions along rows and columns of the chip; the sky
background was then subtracted and spectra extracted.  A low-order
polynomial was fit to the argon lines, and these calibrations applied
to the object spectra. Flux calibration was achieved by dividing the
observed spectra by the spectrum of a nearby mid G-type star and then
multiplying by a model for the absolute flux distribution of the
calibrator.  Residual terrestrial atmospheric features were then
corrected using an early-type star.

\section{Results}
\label{sec:Results}

\subsection{Optical spectra}
\label{sec:Optical-spectra}

The optical spectrum is shown in Fig.~\ref{fig:spectrum}. The spectrum
clearly shows three narrow emission lines --- H$\alpha$\ 
$\lambda$6563, \HeI\ $\lambda$6678, and \HeI\ $\lambda$7065 --- as
well as a broad component to the H$\alpha$\ line which is blue-shifted
with respect to the narrow component. We fit gaussian models to these
lines, using the {\tt specfit} package in {\sc iraf}. After
normalising the spectrum by a low-order polynomial fit to the
continuum, we fit a single gaussian to each of the helium lines and
two gaussians to the H$\alpha$\ line.  The velocities and widths of
the three narrow lines are consistent with being the same; we
therefore constrained them to be equal.  The details of the fit are
shown in Table~\ref{tab:optspec}. The narrow lines show a velocity of
$\sim +378\kms$\ and a width of $\sim 9.5\;$\AA, or $400\kms$, while
the broad component to the H$\alpha$\ line has a very different
velocity ($-310\kms$, or $-688\kms$\ with respect to the narrow lines)
and width ($46\;$\AA, or $2200\kms$).

\begin{table}
\caption{Fit to the optical spectrum.  The spectrum was normalised by a
  low-order polynomial fit to the continuum, and four gaussians were
  fit to the lines, subject to the constraint that the velocity and
  width of the three narrow lines were the same. The table shows the
  resultant velocities, FWHMs, equivalent widths $W_\lambda$\ of the
  gaussians, and the de-reddened line luminosities, assuming a
  reddening of $A_V = 11$\ (Predehl \& Schmitt 1995) and a distance
  of $5.5\;$kpc (Case \& Bhattacharya 1998).}\label{tab:optspec}\addtolength{\tabcolsep}{-1pt}
\begin{tabular}{l r@{\,$\pm$\,}l r@{\,$\pm$\,}l r@{\,$\pm$\,}l c}
Component & \multicolumn{2}{c}{Velocity} & 
\multicolumn{2}{c}{FWHM}   & \multicolumn{2}{c}{$W_\lambda$} & $L_{\rm
  line}$ \\
          & \multicolumn{2}{c}{($\mathrm km\,s^{-1}$)}   & 
\multicolumn{2}{c}{($\mathrm km\,s^{-1}$)} & \multicolumn{2}{c}{(\AA)}
& (${\rm erg\,s^{-1}}$) \\[10pt]
H$\alpha$ broad          & $-$310 & 40 & 2210 & 74 &  50 & 2 & 
\raisebox{-1.5ex}{6.2\ee{35}} \\
H$\alpha$ narrow         &   +378 & 12 &  402 &  8 &  74 & 1 & \\
\HeI\ $\lambda$6678 & \multicolumn{2}{c}{''} & 
 \multicolumn{2}{c}{''} & 3.4 & 0.5 & 1.5\ee{34} \\
\HeI\ $\lambda$7065 & \multicolumn{2}{c}{''} &
 \multicolumn{2}{c}{''} & 4.9 & 0.5 & 2.1\ee{34} \\
\end{tabular}
\end{table}\nocite{ps95}\nocite{cb98}

No stellar features can be seen in the spectrum.  In particular, no
absorption features can be discerned which might sugggest the nature
of the binary companion.  There is no significant variability apparent
between the five spectra. 

Mignani, Caraveo \& Bignami (1997)\nocite{mcb97} obtained a spectrum of
Cir~X-1 using the HST {\em Faint Object Spectrograph} near periastron
(phase 0.18) in June 1995. The line profile of the H$\alpha$\ line in
their spectrum was very different to our 1997 spectrum. They found the
broad component of the gaussian to be centered at approximately its
rest wavelength, while the narrow component was observed at a velocity
of +380\kms. They interpreted the narrow component as arising in an
accretion disc, with the $\sim 400\kms$\ velocity representing the
rotation velocity of the disc material; the absence of the blue
component they interpreted as a phase-dependent shadowing effect.
Under this model, the velocity of the narrow component should shift
with orbital phase, while the broad component remains fixed.

Our spectrum was taken at apastron, and does not show the predicted
shift of the narrow component.  In fact, the velocities we observe for
the narrow component of the H$\alpha$\ line (and the helium lines, not
observed in the HST spectrum) are identical with the HST spectrum,
while the {\em broad} component has shifted in velocity.  This would
seem to indicate that the model proposed by Mignani et al. for the
system is wrong.

\begin{table*}
\begin{minipage}{\textwidth}
\caption{List of observations of Cir X-1, including the archival AAT
  observations, the HST observation (June 1995), and our new AAT
  observation (June 1997). Column 2 gives the instrument/detector
  combination where `B\&C' indicates the Boller \& Chivens
  spectrograph, `RGO' indicates the RGO spectrograph with either the
  long (82~cm) or short (25~cm) camera; columns 3--5 show the
  instrumental resolution, total exposure time, and orbital phase of
  the observation.  The following columns show results of the fit to
  the H$\alpha$\ line (see text for details), with gaussian 1 being
  the red-shifted (narrow) component, and gaussian 2 the blue-shifted
  (broad) component.  The phase was calculated according the ephemeris
  of Stewart et al. (1991), and the fits are shown in
  Fig.~2.}\label{tab:archival-obs} \addtolength{\tabcolsep}{-1.5pt}
\begin{tabular}{llccc r@{$\,\pm\,$}l r@{$\,\pm\,$}l r@{$\,\pm\,$}l
  r@{$\,\pm\,$}l r@{$\,\pm\,$}l r@{$\,\pm\,$}l}
& & & & & \multicolumn{6}{c}{\hrulefill~gaussian 1~\hrulefill} & 
\multicolumn{6}{c}{\hrulefill~gaussian 2~\hrulefill} \\
UT Date         & Instrument     & $\Delta\lambda$ & $t_{\mathrm exp}$
& Phase & \multicolumn{2}{c}{velocity} &
         \multicolumn{2}{c}{FWHM} & \multicolumn{2}{c}{$W_\lambda$} & 
         \multicolumn{2}{c}{velocity} & \multicolumn{2}{c}{FWHM} & 
         \multicolumn{2}{c}{$W_\lambda$} \\
                &                & (\AA)      & (s)       & &
\multicolumn{2}{c}{(${\rm km\,s^{-1}}$)} & 
\multicolumn{2}{c}{(${\rm km\,s^{-1}}$)} & 
\multicolumn{2}{c}{(\AA)} & 
\multicolumn{2}{c}{(${\rm km\,s^{-1}}$)} & 
\multicolumn{2}{c}{(${\rm km\,s^{-1}}$)} & 
\multicolumn{2}{c}{(\AA)} \\[10pt]
1976 May 21     & B \& C/IDS & 15   &  3840 & 0.536 & 370 &  40 & 1050
   &  40 & 380 &  40 &    200 &  60 & 2760 & 340 & 200 & 30 \\
1977 Feb 5      & RGO25/IPCS & 0.8  &   800 & 0.234 & 270 &  20 &  890 
   &  70 & 214 &  13 & $-$830 & 120 & 1200 & 160 &  57 & 10 \\
1977 Sep 4      & RGO25/IPCS & 4.3  &  2000 & 0.947 & 490 &  50 &  640 
   &  50 & 245 &  25 &    160 & 120 & 2080 & 190 & 180 & 25 \\
1978 Apr 29--30 & RGO82/IPCS & 0.45 & 10900 & 0.315 & 410 &  40 &  720 
   &  55 &  52 &   6 & $-$190 &  70 &  630 &  80 &  21 &  6 \\
1978 Aug 11     & RGO25/IPCS & 1.3  &  2000 & 0.522 & 220 & 200 &  700
   & 600 & 180 & 130 & $-$600 &2000 & 1030 & 120 &  30 & 100 \\
1995 Jun 1      & HST/FOS    & 2.2  &  5160 & 0.183 & 330 &  10 &  355
   &  15 &  30 &   2 &  $-$50 &  20 & 1145 &  25 &  75 &  2 \\
1997 Jun 4      & RGO82/TEK  & 5.4  &  9000 & 0.510 & 380 &  12 &  400
   &  10 &  74 &   2 & $-$300 &  50 & 2210 & 100 &  50 &  2 \\
\end{tabular}
\end{minipage}
\end{table*}

\begin{figure}
\centerline{\psfig{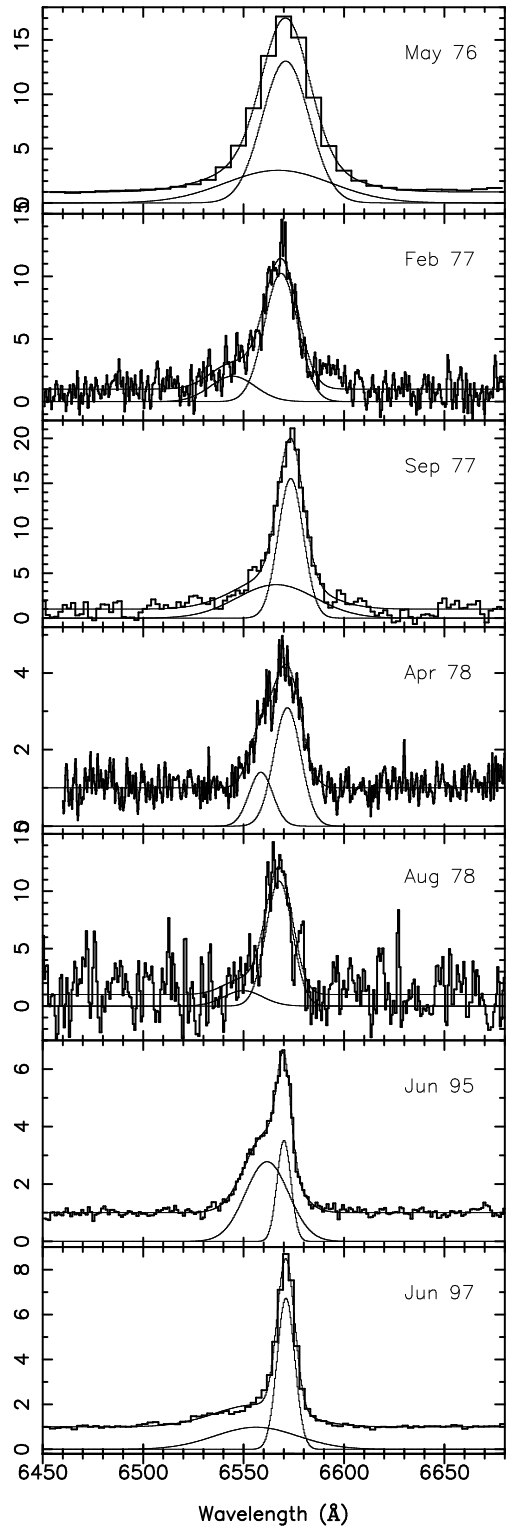}}
\caption{Line profiles of H$\alpha$\ in the archival observations of
  Cir X-1, showing the two gaussians fit to each line and their sum.
  Each spectrum had been binned to twice the spectral resolution, and
  normalised by a polynomial fit to the continuum, so the $y$-axis
  shows the relative flux of the line. The last two spectra show the
  HST observation (June 95) and our new AAT observation (June
  97).}\label{fig:archivalfit} 
\end{figure}

To investigate this, we used data from the AAT archive, which contains
nearly all observations taken with the AAT since its inception.
Cir~X-1 has been observed using the AAT several times by different
observers, with most observations occurring about 20 years ago. We
obtained spectra at a total of five epochs which showed detectable
emission: these are listed in Table~\ref{tab:archival-obs} and shown
in Fig.~\ref{fig:archivalfit}.  The May 1976 data were published by
Whelan \etal\ \shortcite{wmw+77}; the rest of the data are
unpublished.  We obtained the original data from the AAT archive and
reduced them using standard techniques and the {\sc figaro} data
reduction package.  Note that for all these observations, light from
stars 2 and 3 was presumably in the slit.  Because of this, no attempt
has been made to flux calibrate the spectra; instead, each spectrum
was binned to a spectral sampling of two data points per resolution
element, and the resulting spectrum was normalised by a low-order
polynomial fit to the spectrum.

We performed the same double gaussian fit to these spectra as
described in Table~\ref{tab:optspec}, and the results are shown in
Table~\ref{tab:archival-obs}. Several points are immediately clear:
despite the wide range of orbital phases, the broad component is never
observed redward of the narrow component\footnote{Except for the Apr
  78 spectrum, where the width of the red-shifted component is
  marginally smaller than the width of the blue-shifted component;
  however, the two widths are consistent with being identical.}. The
(weighted) mean velocity of the narrow component is $+350\kms$, that
of the broad component $-100\kms$. The scatter of the velocities of
the broad component is much greater than that of the narrow component.
No strong trends can be seen with either date of observation or
orbital phase, with the exception of the equivalent width of the
narrow component, which has been generally decreasing since 1976
(except for the extremely low value in April 1978).

\subsection{Infrared spectra}
\label{sec:Infrared-spectra}

\begin{figure}
\centerline{\psfig{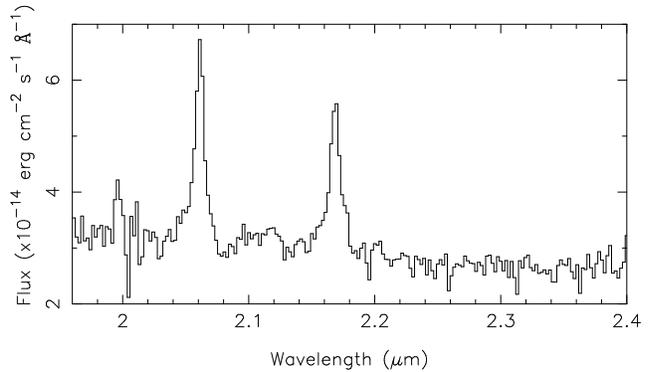}}
\caption{Infrared spectrum of Cir X-1, taken near apastron on 1997
  June 20.}\label{fig:IRspectrum} 
\end{figure}

Infrared spectra were obtained one orbit after the optical spectra.
The mean spectrum is shown in Fig.~\ref{fig:IRspectrum}.  Bright
emission lines of Br~$\gamma$\ and \HeI\ $\lambda$2.058\um\ were
visible in the spectrum. These two lines are generally observed in the
K-band spectra of low- and high-mass X-ray binaries
(\cite{bsc+97,bsct98,csfm99}) and may arise in accretion discs or
larger emitting regions, such as dense winds from the mass donors
and/or ejecta from the binaries.  There also seems to be a hint of
asymmetry in the blue wing of the infrared lines, so we again fit
gaussians to the line profile.  We modelled each emission line as the
sum of two gaussians, constraining the widths and velocities of the
blue and red components to be the same for both lines.  The results
are shown in Table~\ref{tab:IRspec}.

Both lines are well fit by one gaussian at $+450\kms$, a velocity very
similar to the narrow red-shifted optical component, plus a
blue-shifted component at a velocity of $\sim -2000\kms$.

\begin{table}
\caption{Fit to the infrared spectrum. The spectrum was normalised by a
  low-order polynomial fit to the continuum, and four gaussians were
  fit to the lines, subject to the constraint that the velocity and
  widths of the blue (B) and red (R) components were the same for both
  species. The final column shows the de-reddened integrated line
  luminosities, assuming a reddening of $A_V = 11$\ and a distance of
  $5.5\;$kpc.}\label{tab:IRspec}\addtolength{\tabcolsep}{-1pt}
\begin{tabular}{l r@{\,$\pm$\,}l r@{\,$\pm$\,}l r@{\,$\pm$\,}l c}
Component & \multicolumn{2}{c}{Velocity} & 
\multicolumn{2}{c}{FWHM}   & \multicolumn{2}{c}{$W_\lambda$} & $L_{\rm
  line}$ \\
          & \multicolumn{2}{c}{(\kms)}   & 
\multicolumn{2}{c}{(\kms)} & \multicolumn{2}{c}{($\umu{\rm m}$)} &
(${\rm erg\,s^{-1}}$) \\[10pt]
\HI\ Br$\gamma$\ (R)        & 450 & 40 & 1170 & 90 & 88 & 7 &
4.3\ee{34} \\
\HeI\ (R)& \multicolumn{2}{c}{''} &
         \multicolumn{2}{c}{''} & 96 & 7 & 3.1\ee{34} \\
\HI\ Br$\gamma$\ (B)        & $-$1870 & 570 & 1020 & 440 & 12 & 6 \\
\HeI\ (B)& \multicolumn{2}{c}{''} &
                                 \multicolumn{2}{c}{''} & 11 & 5 \\
\end{tabular}
\end{table}

\section{An eccentric low-mass binary model for Cir X-1}
\label{sec:new-model}

Initially, it was suggested that Cir~X-1 is a high mass binary
consisting of a compact star (either a neutron star or black hole) and
an OB supergiant companion star (\cite{wmw+77,mjh+80}). Its X-ray
properties imply a very eccentric binary orbit. The 16.6-day
modulation in the X-ray luminosity is due to orbital variations in the
mass accretion of the compact star. The X-ray and radio bursts occur
when the compact star encounters the dense stellar wind from the
supergiant \cite{ff80}, or when tidal mass transfer is induced
during the periastron passage \cite{hay87}. The Type I X-ray
bursts that were discovered in a brief epoch in the mid-1980s
\cite{tfs86} indicated that the compact star is a neutron star.
Recent observations (\eg\ \cite{snp+91,gla94}) indicate that the
companion star is unlikely to be a massive supergiant star. No Type I
bursts have yet been reported in RXTE observations.

\begin{figure}
\centerline{\psfig{figure=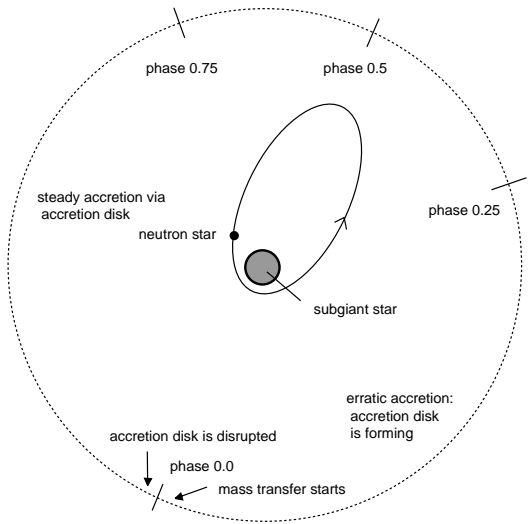,width=7cm,clip=t}}
\caption{Model for Cir X-1. During periastron (phase 0) the
  neutron star passes its closest to the companion (mass $\sim
  3$--$5\protect{\Msolar}$), and the disc is disrupted, resulting in
  large X-ray fluctuations.  After periastron passage, an accretion
  disc begins to form; mass accretion steadies, until after apastron
  (phase 0.5) steady accretion via a disc takes place.  Tidal
  interactions begin to disrupt the disc again as periastron
  approaches.}\label{fig:model}
\end{figure}

Here, we propose a low-mass binary model, where the system consists of
a neutron star orbiting around a subgiant companion star of about 3 to
5\Msolar\ (Fig.~\ref{fig:model}; see also Tennant \& Wu
1998).\nocite{tw98} The orbital eccentricity is $\sim 0.7$--0.9.
During the periastron passage, the companion star overfills its
Roche-lobe, causing a transfer of mass at a super-Eddington rate onto
the neutron star. As Cir X-1 is a X-ray burster, the magnetic field of
the neutron star is relatively weak, and so the accretion flow is
probably quasi-spherical during the periastron passage.  Because of
the large radiative pressure of emission from the neutron star, there
must also be a strong (anisotropic) matter outflow. The inflow/outflow
geometry may be related to the larger scale jets observed from the
system, with symmetrical collimated outflows along some preferred
axis.

After the periastron passage, the companion star is detached from its
critical Roche-surface, and mass transfer ceases. Accretion continues
as the neutron star captures the residual matter in its Roche-lobe. An
accretion disc is gradually formed. The disc is geometrically thick,
because the accretion rate is sufficiently high and is close to the
Eddington limit. The accretion becomes more steady when disc accretion
takes over from the less stable super-Eddington quasi-spherical
accretion.  Accretion via a disc continues to operate till the next
periastron passage, where the accretion disc is disrupted by tidal
interaction.  When the periastron passage proceeds, the companion star
overfills its Roche-lobe again, leading to another cycle of
super-Eddington mass transfer, formation of the accretion disc, steady
accretion via the disc and disruption of the accretion disc.

The periodic X-ray light curve of Cir X-1 has changed dramatically
since its discovery in the 70s (\eg\ 
\cite{khbs76,tkm+89,sblm96}).  This indicates that the accretion
mode has varied substantially in the last 25 years. Despite the fact
that the H$\alpha$\ lines in the optical spectra obtained in the 70s
and 90s show certain similarities, such as an asymmetric profile with
a broad blue wing, it is inappropriate to assume that the optical
spectra are the same at the same orbital phases in the 70s and the
90s. The FWHM of the red (narrow) component of the H$\alpha$\ line
seems to decrease from $\approx 700\kms$\ in the 70s to $\approx
380\kms$\ in the 90s. The central wavelength, however, does not show
significant velocity variation, and its average value is
$350\,\pm\,80\kms$.

Our interpretation is that the red (narrow) component is emitted from
irradiatively-heated matter which is relatively stationary with
respect to the centre of mass of the system and has a low velocity
dispersion. One of the natural choices is the heated surface of the
companion star, on which the projected velocity of the matter is much
less than the flow velocity near the neutron star.\footnote{Note that,
  if the narrow emission is arising from the heated face of the
  secondary, the equivalent width should vary in a smooth fashion
  through the binary orbit, with a maximum at phase 0.5; such a
  pattern is not seen in the data in Table~\ref{tab:archival-obs}.
  However, we argue that the mass transfer rate has varied so much
  over the 21~y spanned by the observations that it is misleading to
  try to compare the equivalent widths.  Indeed, if we consider three
  groups of spectra near to each other in time (1976--1977, 1978, and
  1995--1997) there is a (slight) tendency for the equivalent width to
  be highest at phase 0.5.  Such a relation certainly needs to be confirmed.}

The velocity of the sharp red component of the \HI\ Br$\gamma$\ and
the \HeI\ $\lambda 2.058\um$\ lines is $450\,\pm\, 40\kms$, which is
consistent with that for the optical lines. Thus, the lines must also
be emitted from a region relatively stationary to the centre-of-mass
of the binary. The width of the IR lines, however, is much larger than
the width of the optical lines. It is therefore unlikely that the
emission site is the surface of the companion. One possibility is that
these IR lines are emitted from a heated dust shell -- the residue of
ejecta from previous epochs.

The observations accumulated since the 70s show that the broad
component is blueward of the narrow component, except (possibly) in
the 1978 April observation. The line widths were generally above
1000\kms, and it was even larger than 2000\kms\ in the 1976 and 1997
observations. Although the line centre velocities appear to show
large variations (from $-600\kms$\ to $200\kms$), they show no obvious
correlation with orbital phase. 

We suspect that the broad component is related to the high velocity
flow from regions near the neutron star. X-ray observations have shown
that the rate of mass accretion onto the neutron star is above the
Eddington limit (e.g.\ Inoue 1989).\nocite{ino89} It is therefore
possible that an outflow is driven by the Eddington luminosity. As we
only see the blueward component, the lines are probably emitted from
the optically thick outflowing matter, of which the red-shifted
components are heavily absorbed or obscured. The changes in the line
centre velocities and widths deduced from the fits are probably the
result of the changes in optical depth over the different
observational epochs.

%(see simulations in Fig.~\ref{fig:sim-main})

\begin{figure}
     \centerline{\psfig{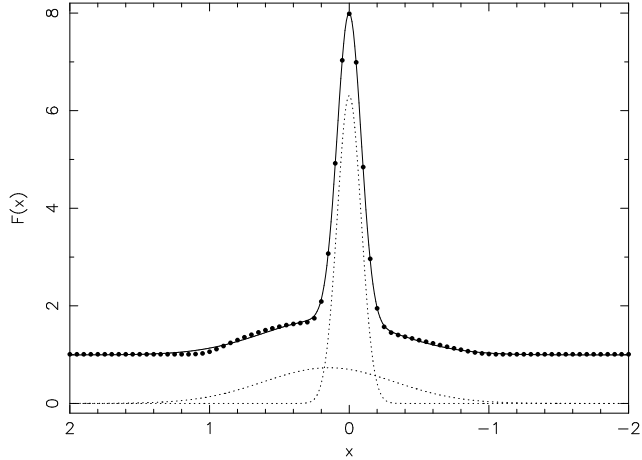}}
     \caption{Line profile of emission from an expanding envelope,
       with parameters as described in the text. Positive x
       corresponds to the blue-shifted frequency; negative x
       corresponds to the red-shifted frequency. The filled circles
       are the simulated data. The solid line is a fitted model with a
       constant and two gaussians. The two gaussians are represented
       by the dotted lines.  The ratio of the widths of the broad and
       narrow gaussians is 5.5, and the ratio of the width of the
       broad gaussian to the separation of the two gaussian peaks is
       3.22.}\label{fig:sim-main}
\end{figure}

The quality of our data precludes us performing a fit of this model to
the line profile; rather, we present it as a possible explanation for
the observed asymmetric line profile.  We have constructed a simple
toy model to demonstrate the asymmetric line profile that might be
observed from such an outflow. In our model the broad line emission
region is a spherically symmetric extended envelope, with outer and
inner radii $R_{\rm out}$\ and $R_{\rm in}$\ respectively. The envelope
encloses an opaque sphere with radius equal to $R_{\rm in}$. We have
considered various functional forms for the velocity, emissivity and
absorption coefficient distribution, and have found that models with a
power-law velocity profile generally can produce a broad line profile
similar to the observations (see Appendix).

In Fig.~\ref{fig:sim-main}, we show a simulated a blue-shifted broad
line from an expanding envelope with outer and inner radii, $R_{\rm
  out} = 1$\ and $R_{\rm in} = 0.3$\ respectively. The radial
expanding speed $V$\ has a power-law profile with an index of +1, and
it is normalised such that its value is 1 at $R_{\rm out}$. The
normalised local Doppler width of the line is 0.1. The emissivity is
uniform within the emitting region, with $\eta_c = \eta_l =1$. The
continuum and line absorption coefficients are uniform, and their
ratio $\chi_l/\chi_c = 0.35$. The total effective absorption optical
depth $\tau$\ is such that $(\chi_c + \chi_l) R_{\rm out} = 1$.

As the narrow line is emitted from the heated companion star, we
simply assume it is a gaussian with a width of 0.123 and a norm of
4.0. This will give an approximate ratio of 2/3 for the equivalent
widths of the simulated broad and narrow line components.

As shown, the resultant line profiles resembles the H$\alpha$\ line
observed in our June 1997 data. If we fit the simulated line with two
gaussians and a constant, the relative widths and shifts of the two
gaussian components are in good agreement with the fitted parameters
of the June 1997 data (Fig.~\ref{fig:sim-main}).

The blue components of the \HI\ Br$\gamma$\ and the \HeI\ $\lambda
2.058 \um$\ lines have a velocity shift of $-1870\,\pm\, 570\kms$\ and
a width of $1020\,\pm\, 440\kms$.  If we interpret the blue component
of the IR lines as being emitted from the vicinity of the neutron
star, the difference in the velocities between the optical and IR
lines is due to optical depth effects and/or the uncertainties in
determining the central velocities of the lines with the spectral
resolution of the IR spectrum.

Murdin \etal\ \shortcite{mjh+80} assumed a 1\Msolar\ compact star, and
by comparing the radius of various classes of stars with the
separation of the two component star at periastron, they estimated the
mass of the companion star and the orbital eccentricity for which
tidally induced mass transfer can occur during the periastron passage.
In our model, the mass transfer is driven by the overflow of the
mass-donor star's Roche-lobe instead of the stellar wind from the
star, and it therefore requires the companion star to fill its
Roche-lobe at periastron, \ie\ the star's radius $R_2$\ to be equal to
or larger than its `instantaneous' critical Roche-lobe surface. It is
beyond the scope of this paper to derive an exact phase-dependent
effective Roche-lobe surface.  However, as a first approximation, if
we simply take the expression of the Roche-lobe radius for circular
systems (see Kopal 1959)\nocite{kop59} and replace the orbital
separation term by the separation between the two stars of Cir X-1 at
periastron, then we can obtain an effective Roche-lobe radius at
periastron (phase~0):
\begin{equation}
     R_{_{\mathrm L}} \approx a(1-e) [ 0.38 + 0.2 \log q ]
\end{equation}
where the mass ratio $q = M_2/M_1$\ and $a$\ is the semi-major axis.
Using this approximate expression for $R_{_{\mathrm L}}$, we can
constrain the appropriate parameter space ($e$; $M_1$, $M_2$) where
Roche-lobe overflow occurs during the periastron passage for various
types of mass-donor stars.

\begin{figure}
\centerline{\psfig{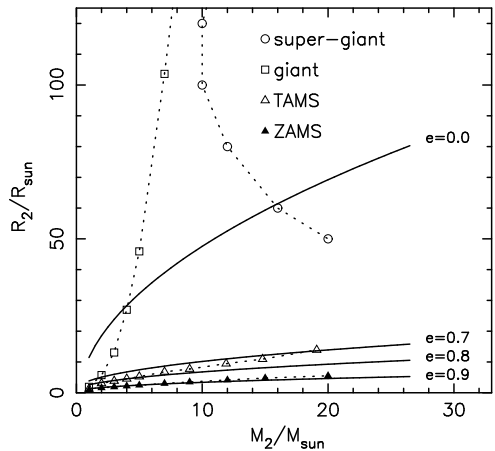}}
\caption{The radius of the Roche-lobe at periastron as a function of
  the mass of the companion star, calculated from an eccentric system
  with a 16.6~d period and a 1.4\Msolar\ neutron star primary. For
  Roche-lobe overflow to occur, the size of the secondary must exceed
  or equal the volume enclosed by the instantaneous critical
  Roche-lobe surface. We also show the size of the star for various
  masses at the zero-age main-sequence (ZAMS), terminated
  main-sequence (TAMS), giant and supergiant stages (indicated with
  filled triangles, open triangles, open squares and open circle
  respectively). The radii of the stars are derived from the
  evolutionary models of Maeder \& Meynet (1988), with the ZAMS, TAMS
  and giant stages corresponding to Point 1, 7 and 14 in the
  evolutionary tracks. The mass-radius relation of the super-giant
  stars, which is not as well-defined, is adopted from Lang
  (1991).}\label{fig:RLradius}
\end{figure}\nocite{lan91}\nocite{mm88}

In Fig.~\ref{fig:RLradius} we show the mass-radius relation for
supergiant, giant and main-sequence stars, together with $R_{_{\mathrm
    L}}$\ for eccentric systems with a 16.6 day period and a
1.4\Msolar\ neutron star primary. If the companion star is a main
sequence star, its mass is about 10\Msolar\ and the orbital
eccentricity $e > 0.9$\ in order to fill its Roche-lobe at periastron.
A sub-giant star with $M_2 \approx 3$--$5\Msolar$\ can easily fill its
Roche-lobe for an orbital eccentricity of about 0.7--0.9.

The argument for a low mass system is generally consistent with the
combined results of our spectroscopic observations and the study on
the effects of supernova kicks on neutron star binaries. A simulation
study \cite{bp95b} showed that the eccentricity of HMXBs (with $M_2
\sim 10\Msolar$) and LMXBs (with $M_2 \sim 1$--$5\Msolar$) resulting
from supernova explosions with isotropic kicks will be about 0.5
($\sigma \approx 0.2$) and 0.8 ($\sigma \approx 0.05$) respectively if
the final orbital period is about 16 days. The corresponding
centre-of-mass velocities are 50 ($\sigma \approx 12$) and 150
($\sigma \approx 100$)\kms. If we accept the interpretation that the
narrow line components originate from the heated companion star, the
line-of-sight component of the centre-of-mass velocity of the system
is about 350\kms, or 430\kms after applying a correction for Galactic
rotation (using the rotation curve of Clemens 1985)\nocite{cle85}.
This correction was made assuming that the position of G321.9$-$0.3
was the birthplace of Cir~X-1, but this makes little difference to the
resulting correction.  By combining with the transverse velocity
($\sim 300$--$400\kms$) inferred from the age and the distance of the
supernova remnant G321.9$-$0.3 \cite{cpc75} we obtain an estimated
centre-of-mass velocity of $\sim 540\kms$.  This velocity could be
substantially higher, as it depends on the (very uncertain) age
estimate of the supernova remnant; transverse velocities as high as
1600\kms are possible \cite{schn93}. A velocity of $\sim 540\kms$\ is
about 3$\sigma$\ from the mean centre-of-mass velocities obtained from
the simulation study by Brandt \& Podsiadlowski \shortcite{bp95b}.
Such high velocities are not at all consistent with the simulations
for high mass systems: even the low velocity is about 13$\sigma$\ 
above the mean velocity predicted for a high-mass system.

The implied velocity for Cir~X-1 makes it one of the fastest moving
binaries known: radial velocities of $\sim 300\kms$\ have been
measured for a handful of other LMXBs, \eg\ 4U~1556$-$605
\cite{mpmb89}, GX~349+02 \cite{pa91}, 4U~1957+11 \cite{mtb78}; see
Johnston \shortcite{joh92} for a discussion. Such velocities
strengthen the identification of Cir~X-1 as a low-mass system, as the
fastest moving HMXBs have velocities $v < 100\kms$.  If the transverse
velocity is similar to the radial velocity, this implies a proper
motion of 15~${\mathrm mas}\,{\mathrm yr}^{-1}$; in an ongoing project
to measure the proper motion of the radio counterpart to Cir~X-1, we
can measure a relative positional accuracy of around 40~mas, so we
should measure the proper motion in a couple of years.

An alternative simulation by Shirey \shortcite{shi98} also concluded
that Cir~X-1 is a low-mass system.  However, his simulation requires
the system to be old ($\sim 10^7$~y), which seems difficult to
reconcile with the observations that Cir~X-1 is associated with the
supernova remnant G321.9$-$0.3.
 
The RXTE observations show that the current X-ray luminosity is quite
steady in the orbital phases around 0.5--1.0. The cycle-to-cycle
variations are insignificant in comparison with the those at the
earlier orbital phases. We interpret the steady accretion as due to
the presence of an accretion disc. The erratic cycle-to-cycle
luminosity variations are the consequence of the absence of the disc
acting as a buffer; cf. the rms fluctuations in the RXTE data at
phases 0.15 and 0.5 \cite{sblm96}. Moreover, we do not see
double-horn features in the H$\alpha$\ line, such as those often
observed in cataclysmic variables (\eg\ \cite{hm86}) or black hole
binaries (\eg\ \cite{swhw98}). This, together with the
near-Eddington luminosity of the X-rays during the disc-accretion
phase, implies that the disc is non-Keplerian and it is both
geometrically and optically thick. As the matter is not tightly bound
by the gravity of the neutron star, the large variation in the tidal
force when the neutron star approaches periastron could easily disrupt
the disc, causing a short temporal decrease in the accretion
luminosity of the system.

What constraints can we put on the age of the system?  In our model,
mass transfer at periastron does not begin until the star has evolved
from the main-sequence towards the giant branch. For a 3--5\Msolar\ 
star, the subgiant stage may last $\sim 5\ee{5}$~y. If Cir X-1 is
associated with the nearby supernova remnant G321$-$0.3, then the
implied age of the system as an eccentric neutron star binary is $<
10^5$~y. The Eddington limit for the rate of mass transfer onto a
neutron star is $\sim 3\ee{8}\Msolar\;{\mathrm y}^{-1}$, giving a mass
transfer time scale of $3\ee{7}$~y. Since the companion star in Cir
X-1 can lose mass at a rate greater than this as an average in each
orbital cycle, the life-span of the system as an X-ray binary is
correspondingly shorter, but still comfortably above either the
evolution time scale of the companion star or the age of the supernova
remnant.

\section{Conclusions}
\label{sec:Conclusions}

We have detected strong optical and infrared emission lines in the
spectrum of Cir~X-1; the H$\alpha$\ and infrared lines are asymmetric,
with a narrow component at a velocity of $\sim +350\kms$\ and a
broader, blue-shifted component.  Archival optical observations show
that an asymmetric H$\alpha$\ emission line has been in evidence for
the past twenty years, although the shape of the line has changed
significantly. The narrow component is always seen redwards of the
broad component, at a velocity around 200--$400\kms$; the broad
component has a much larger scatter of velocities.  These properties
are not consistent with the interpretation of Mignani \etal\ 
\shortcite{mcb97}, who suggested that the narrow component arises in
an accretion disc, with the absence of a red component due to a
phase-dependent shadowing effect.

Instead, we suggest that the narrow component arises from the heated
surface of the companion star, which is a subgiant of about
3--5\Msolar, while the broad component arises in an optically thick,
high velocity outflow driven by super-Eddington accretion onto the
neutron star. During periastron passage, the companion star overfills
its Roche-lobe, causing a transfer of mass at a super-Eddington rate,
which in turn drives a strong matter outflow. After periastron, mass
transfer from the companion ceases, but accretion continues at a
near-Eddington rate as the neutron star captures the residual matter
in its Roche-lobe. An accretion disc gradually forms, which is
non-Keplerian and geometrically thick. This change between
quasi-spherical and disc accretion causes the change between strong
X-ray variability near phase 0 and steady accretion between phases
0.5--1.0. Thus our observations are in general consistent with a
low-mass binary model for the system.  In this model, the velocity of
the narrow component reflects the space velocity of the binary; the
implied radial velocity (+430\kms) makes Cir~X-1 one of the fastest
moving binaries known.

\section*{Acknowledgments}
\label{sec:Acknowledgments}

We thank Allyn Tennant, Peter Wood, Lex Kaper, and Thomas Tauris for
useful discussions; we also thank Peter for assistance with our IR
observation.  RPF was supported during the period of this research
initially by ASTRON grant 781-76-017, and subsequently by EC Marie
Curie Fellowship ERBFMBICT 972436. KW is supported by ARC through an
Australian Research Fellowship. This research was based in part on
observations made with the NASA/ESA Hubble Space Telescope, obtained
from the data archive at the Space Telescope Science Institute. STScI
is operated by the Association of Universities for Research in
Astronomy, Inc. under NASA contract NAS 5-26555.

\appendix
\section{Line profile simulations}
\label{sec:Line-prof-simul}

The transfer equation that determines the intensity of the 
radiation is 
\begin{equation}
\biggl[ {1 \over c}{\partial \over {\partial t}} + 
   ({\hat n} \cdot \nabla) +  \chi (\nu, {\vec r}) 
   \biggr] I(\nu, {\vec r})\ =\ \eta (\nu, {\vec r}), 
\end{equation}
where $I(\nu, {\vec r})$\ is the intensity, ${\hat n}$\ is the unit 
vector in the direction of the ray propagation, $\eta (\nu, {\vec r})$\ 
is the emissivity and $\chi (\nu, {\vec r})$\ is the extinction 
coefficient. We assume a (quasi-)stationary state, such that the term 
${\partial/ {\partial t}}$\ can be set to zero. In a general situation, 
the extinction coefficient consists of an absorption term and a 
scattering term. Therefore, one needs to solve an integro-differential 
equation to determine the radiation field and then the emerging 
intensity. However, if we can neglect scatterings, the situation is 
greatly simplified and a simple formal solution can be obtained. 

\begin{figure}
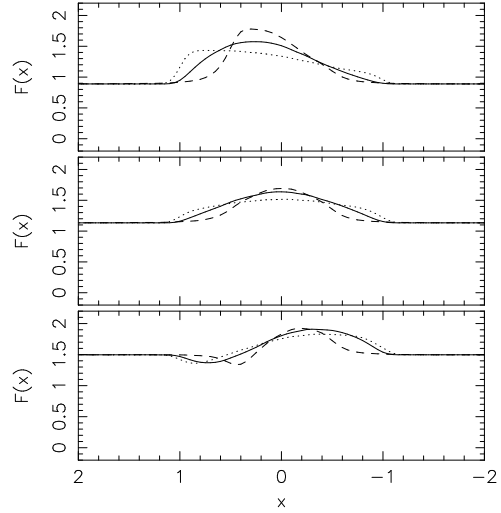

\centerline{\psfig{figure=jfw_figa1.ps,width=6.5cm,clip=t}}
\centerline{\psfig{figure=jfw_figa2.ps,width=6.5cm,clip=t}}
\centerline{\psfig{figure=jfw_figa3.ps,width=6.5cm,clip=t}}
\caption{Simulated broad line components from expanding envelopes for
  a power-law radial velocity profile with various power-law indices.
  From top to bottom, $\chi_l/\chi_c = 0.1$, 1 and 10. The other
  parameters are: $\eta_l =1$, $\eta_c =1$, $R_{\mathrm out} = 1$,
  $R_{\mathrm in} = 0.1$, $\delta = 0.1$ and $(\chi_l +
  \chi_c)R_{\mathrm out} = 1$.  The velocity normalisations
  corresponding to the three cases with the power-law indices $+1$\ 
  (solid line), 0 (dotted line) and $-1$\ (dashed line) are
  $V(R_{\mathrm out}) =$ 1, 1 and 0.45 respectively.  }\label{fig:A1}
\end{figure}

In our model, we assume that the radiation is emitted from an envelope
within which matter is moving with velocity $\vec v ({\vec r})$.
Because the emitters are in relative motion to the observer, there are
velocity shifts in the frequencies of the radiation they emit. There
is also a velocity spread among the emitters, and so the
frequency-dependence of the local line absorption and emission
coefficients are modified.

For mathematical convenience we consider a dimensionless frequency 
$x (\equiv c(\nu - \nu_o)/\nu_o v_{*})$, and a dimensionless velocity 
${\vec V} ({\vec r}) (\equiv {\vec v}({\vec r})/ v_{*})$, where $\nu_o$\
is the frequency in the rest frame, $v_{*}$\ is a velocity normalisation 
and $c$\ is the speed of light. The emissivity and the absorption 
coefficients are decomposed into a continuum and a line component. As 
a first approximation, we assume that the continuum component does not 
vary significantly with frequency near the line frequencies, and hence 
we have   
\begin{equation}
 \eta (x, {\vec r}) \ = \  \eta_{c}({\vec r})  
           + \eta_{l}({\vec r})\phi(x, {\vec r}, {\vec V}) \ ; 
\end{equation}
\begin{equation}
 \chi (x, {\vec r}) \ = \ \chi_{c}({\vec r})  
           + \chi_{l}({\vec r})\phi(x, {\vec r}, {\vec V}) \ . 
\end{equation}
In the above expression, $\phi(x, {\vec r}, {\vec V})$\ is the line 
profile, given by  
\begin{equation}
\phi(x, {\vec r}, {\vec V})\ = 
   \ {1 \over {\delta ({\vec r})  \sqrt{\pi}  }}\  
   \exp \{-[ x - ({\hat n}\cdot {\vec V})]^2/\delta^2({\vec r})\}\ ,
\end{equation}
where $\delta({\vec r})$\ is the normalised local Doppler width.  

As the rays that reach a distance observer are parallel, we can assign
each of them an impact parameter $p$\ with respect to the centre of
the emitting envelope. The flux at a frequency $x$\ seen by the
observer is the sum of all rays, i.e.\ 
\begin{equation}
 F(x) \ =\  \int d\varphi \int dp \ p \ I^*(x, p) \, 
\end{equation}
where $I^*(x,p)$\ is the intensity of the ray emerging from 
non-obscured part of the emitting envelope ${\Re}$, and 
$\int d\varphi \int dp~p$\ is the projection area of the expanding 
envelope on the sky plane. If we neglect scatterings, 
$I^*(x,p)$\ is the formal solution to the radiative 
transfer equation, and it is    
\begin{eqnarray} 
 I^*(x, p) & = & I_o(x,p)\ \exp\{ -\tau(x,p,z_{\mathrm min}(p))\} 
   \\ 
   & + & \int_{z_{\mathrm min}(p)}^{z_{\mathrm max}(p)} dz\  
   \big[ \eta_{c} + \eta_{l}\phi \big]\  
 \exp\{ -\tau(x,p,z(p))\} \nonumber 
\end{eqnarray}  
(e.g.\ \cite{mih78,kyf88}), where $dz$\ is a path element along
the ray propagation, and $I_o(x,p)$\ is the intensity of the
background radiation. The optical depth $\tau(x,p,z(p))$\ is
\begin{equation}
  \tau(x,p,z(p)) \ =\  \int_{z(p)}^{z_{\mathrm max}(p)} dz\  
   \big[ \chi_{c} + \chi_{l}\phi \big]\   . 
\end{equation}
If there is no strong background radiation behind the emitting envelope 
(i.e.\  $I_o(x,p) \approx 0$), then the observed flux is simply    
\begin{eqnarray}
  \lefteqn{F(x) = \int d\varphi \int dp \,p 
  \int_{z_{\mathrm min}(p)}^{z_{\mathrm max}(p)} dz 
   \big[ \eta_{c} +  \eta_{l}\phi \big] } \hspace{3cm} \nonumber \\ 
    && \times \exp\{ -\tau(x,p,z(p))\}.  
\end{eqnarray}

We consider a hybrid method to calculate the line flux $F(x)$. The 
optical depth $\tau (x,p,z(p))$\ is determined by a direction integration 
of the absorption coefficient along the ray propagation ${\hat n} z$\
(for fixed $p$). Instead of carrying out a direct integration in the 
3-D space $(\varphi,p,z)$, we use a Monte-Carlo method to determine
$F(x)$. 

In the Monte-Carlo calculation, we consider each emitter as a discrete
source which emits a packet of line and continuum radiation spanning a
certain frequency range. The packet is absorbed along the path of
propagation subject to the optical depth. The strength of the emission
from each emitter (i.e.\ $\eta_{c}$\ and $\eta_l$) is weighted
according to a distribution profile in the envelope.  We consider
50000 emitters in each simulation. The observed flux is the sum of the
contribution of all emitters in the non-obscured region $\Re$.

This quick-fix method allows us to handle complicated geometry due to 
obscuration easily. It also enables us to survey various functional forms 
for the distribution profiles of the velocity, emissivity and absorption 
coefficients and obtain qualitative features of the line emission with 
sensible computational time.  

In Fig.~\ref{fig:A1} we show examples of line profiles for three cases
in which the radial velocity distribution is a power-law function
(with indices $-1$, 0 and $+1$\ respectively). In these cases, if the
line absorption is weaker than the continuum absorption, i.e.\ $\chi_l
< \chi_c$, an asymmetric blue-shifted profile can be obtained.  If
$\chi_l \approx \chi_c$, the line is symmetric with no obvious shift
of the line centre frequency.  If $\chi_l > \chi_c$, a P-Cygni profile
will be obtained.

%\bibliographystyle{aabib}
%\bibliography{strings,refs}

\label{lastpage}

\end{document}